**HTS Potential and Needs for Future Accelerator Magnets**

L. Bottura, B. Bordini

**Abstract**

HTS has the potential of a game changer for many applications of superconductivity, not last in the field of particle accelerators and detectors. This paper explores the potential of HTS, with a focus on REBCO-coated conductors, in relation to the evolving demands of superconducting magnets for accelerators. HTS already have a spectacular current carrying ability at high field, demonstrated and available on relevant lengths. Recent advances in non-conventional winding techniques for solenoids, in particular non-insulated windings, have shown that it is possible to reach engineering current densities in the coil exceeding by far those of LTS . This approach seems to offer an extended field reach, as well as solutions to the challenges associated with magnet mechanics, quench management and cost. Most important, beyond the ability to reach a field range higher than what is possible with LTS, HTS offers an extended range of operating temperature, with large margin. This can be exploited to obtain higher availability and better cryogenic efficiency, a must for the future of sustainable large scale research infrastructures such as particle accelerators.

**Introduction**

Superconducting magnets are one of the key technologies of present and future accelerators and colliders for High Energy Physics (HEP). The long thread of colliders built with Nb-Ti superconducting magnets started with the Tevatron in 1983 [WIL-1978], continued with HERA in 1991 [MEI-1991], and RHIC in 2000 [ANE-2003], reaching state-of-the-art with the completion of the LHC in 2008 [EVA-2008]. This thread is likely to continue with the HEP experiments of the future.

It is the case of the High-Luminosity LHC upgrade (HL-LHC) [BRU-2015], presently being built at CERN and collaborating laboratories. The HL-LHC project, initiated in 2015, aims at improving the precision of the LHC physics by a factor 3 or more. Its cornerstone in terms of magnet technology are the quadrupoles of the low-beta triplets at point 1 and 5 [TOD-2021], as well as the 11 T dipoles [BOR-2019] for the dispersion suppressor. Both are built using a high-$J_C$ grade of $Nb_3Sn$, and besides their function in the HL-LHC, they are intended as a technology demonstration of the first use ever of this superconducting material in a running accelerator. Indeed, one of the aims of HL-LHC from its outset was to show that $Nb_3Sn$ is a viable option for the next step machine in the post-LHC era.

Many alternatives are considered for a next step collider, all still at study level. Confining ourselves to circular colliders, a first relevant example is the Future Circular Collider integrated design study (FCC) hosted by CERN [ABA-2019-1], consisting of a lepton collider FCC-ee, up to 375 GeV center-of-mass energy, followed by a hadron collider FCC-hh in the same tunnel, with 100 TeV center-of-mass energy. A second example is a Muon Collider (MC) with center-of-mass energy of 10 TeV, studied in the framework of the International Muon Collider Collaboration (IMCC) also hosted by CERN [SCH-2021]. These studies represent to date the major European effort in the field. Similar studies are on-going outside Europe, such as the Chinese IHEP proposal for a sequence starting with a Circular Electron Positron Collider (CEPC), followed by a Super proton-proton Collider (SppC) with a center-of-mass energy ranging from 70 to 120 TeV, all within the same tunnel [CEP-2018, SUF-2016]. It is likely that the US will soon join in the efforts for a circular collider, taking a role in the study of options for a muon collider with center-of-mass energy in the range of 3 to 10 TeV [PPP-2023, CHA-2024]

As in the past [BOT-2021], the common denominator of all the above studies is the need for better, stronger magnets. This generally implies an extended field reach, exceeding both the present limits of Nb-Ti (i.e. the 8.33 T of the LHC) [ROS-2003], as well as $Nb_3Sn$ (i.e. the 12 T of HL-LHC) [TOD-2021, BOR-2019]. To give representative examples, the FCC-hh under study at CERN foresees dipole magnets with a 50 mm free bore and nominal field originally set at 16 T [ABA-2019], but now adjusted in the range of 14 T [SEG-2025] to 20 T. Similar values are considered for the SppC at IHEP, with a staged approach that foresees a first step with 12 T dipoles with a free bore of 50 mm [WAN-2016], to be followed by an energy upgrade with 20 T dipoles and 50 mm free bore [XUQ-2015]. The above field progression appears to be in line with the history of hadron colliders. By contrast, new accelerator concepts such as the muon collider pose additional and significant challenges on the magnetic system [BOT-2024-1]. Accelerator dipole and quadrupole magnets in the range of 10 to 20 T peak field [NOV-2025] are not the only *breed* required for such a machine. A muon collider also requires solenoids with a wide range of specifications: from very high field solenoids, reaching up to 20 T with a large bore of around 1.5 meters [BOT-2024-2], to ultra-high field solenoids exceeding 40 T with a more modest bore of 50 mm [BOR-2024]. Moreover, magnets in various locations within a muon collider are subject to substantial heat deposition (several watts per meter of magnet length) and significant radiation doses (several tens of MGy) [CAL-2022].

In this paper we start with a review of the cost scaling and constraints for any of the above projects, beyond the common aspiration to extend the physics reach. We then elaborate on how HTS magnet technology, yet to be developed, can respond to the drivers and meet the constraints, not only from the point of view of performance, but also in view of affordable investment and sustainable operation. We finally briefly outline the main challenges for HTS materials in accelerator magnets. Challenges are partly in common with other fields of science and societal applications, which provides a strong motivation for further development and implementation of these extraordinary materials in accelerators.

**Cost projections of future colliders**

Future colliders at the energy frontier are likely to be significantly larger, or more performant, or both, when compared to the state of the art. Audacious examples quoted above are the FCC [ABA-2019-1], or the CEPC/SppC [CEPC-2018, SUF-2016], with tunnel lengths in the range of 100 km and dipole magnets in the range of 14 to 20 T. Cost, both the initial investment, or capital expenditure (CAPEX) as well as the operation expenditure

(OPEX), is an obvious concern. Fortunately, past realizations have shown that technological advances tend to progress hand-in-hand with the demands of physics, leading to a reduction in the specific cost C of the technical infrastructure for accelerators, as illustrated in Fig. 1., from [LEB-2017]. Note that we plot there the specific capital expenditure per parton center-of-mass energy $E_{pCOM}$, assuming a factor seven for the equivalence between protons and leptons [BAR-1997]. This representation is better to compare the colliders of different particles, nature and technology that span over several decades. The specific cost appears to scale well with the parton collision energy, following a power law:

$$C[MEUR/GeV] = \frac{700}{E_{pCOM}^{0.72}[GeV]} \qquad (1)$$

This scaling predicts that the specific cost for the next step at the energy frontier, in the range of 10 TeV pCOM energy [ESG-2020, PPP-2023], should reach a target of 1 MEUR/GeV, corresponding to a cost of the technical infrastructure of 10 BEUR. In fact, this value is close to the cost estimate of the only magnet system of the FCC-hh, approximately 9.4 BEUR [ABA-2019-2]. As the contribution of other technical systems to the total cost is not negligible, this already indicates the need for an evolution towards a magnet technology with lower cost per unit length.

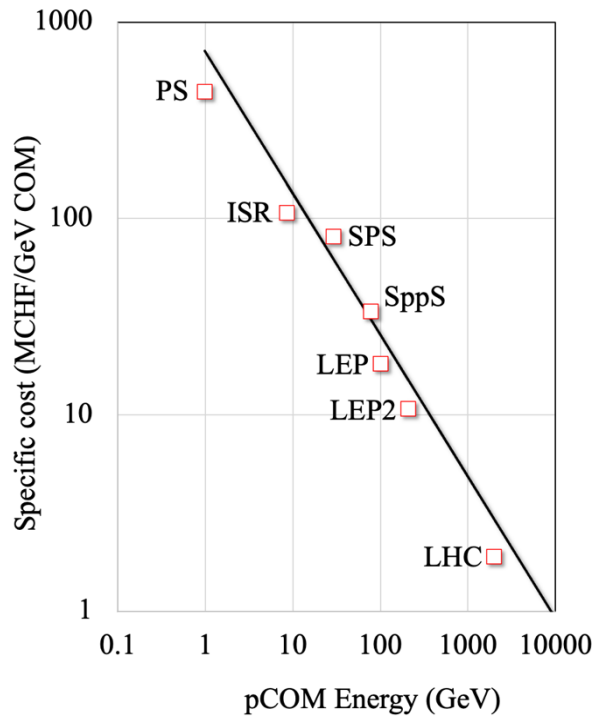

Figure 1. Specific cost of technical systems, referred to the parton center of mass energy, for all colliders built at CERN. Reproduced from [LEB-2017].

In addition, even if we allocated the whole technical infrastructure cost to the magnet systems, for an FCC-hh with a tunnel length of 90.7 km and a filling factor of 82 % the above target corresponds to a maximum cost per unit length of approximately 130 kEUR/m. To put this figure in perspective, the present cost of $Nb_3Sn$ HL-LHC magnets is estimated at 400 kEUR/m[1], i.e. about three times larger than targeted for the FCC-hh. Alternative concepts such as the muon collider, whose ring length is approximately 10 km [REF], could afford higher specific costs, in the range of 300 to 400 kEUR/m, closer to the present cost of high field accelerator magnets.

It is hence clear that a straight extrapolation of present technology is not likely to be enough to enable the next step in particle physics at the energy frontier. In fact, as mentioned above, the steady decrease of specific cost of colliders was mainly achieved through advances in technology rather than optimization of processes. As we will expand later, we believe that this can be achieved only by developing a HTS magnet technology suitable to accelerators.

**Cryogenics for sustainable science**

A circular collider at the energy frontier will make extensive use of superconducting magnets, which require cooling at cryogenic temperatures. Given the size of a next step collider, cryogenic will represent one of the main users of electricity. Also, the use of helium has been associated in the last years with a significant risk of availability and price volatility, both needing appropriate consideration.

To set on relevant orders of magnitude, the FCC-hh conceptual design for a hadron collider with 100 TeV center-of-mass energy projects an electricity consumption of about 250 MW for cryogenics (after optimization) [BEN-2022], i.e. about six times higher than the present electricity consumption of the cryogenic plant for LHC. The helium inventory of about 880 tons [ABA-2019-2], is nearly an order of magnitude more than the present inventory in the LHC. Even with a perfected closed cycle system such as the LHC, typical helium losses are from 5 to 10 % per year of operation [CLA-2015, BRO-2015]. In the case of the FCC-hh this would result in yearly losses of 50 to 100 tons per year, a considerable amount of helium, associated cost and availability risk.

While the figures above are not beyond the limit of possibility, present yearly production of helium is over 25,000 tons/year [MOH-2014, PRO-2022], priority should be set in reducing both electrical consumption and inventory. Besides the ethical commandment to reduce footprint, this is important also to make the next step collider an affordable and reliable instrument, whose operation does not depend critically on fluctuations of electricity prices and availability of cryogens for basic science.

Restricting our attention to helium as the cryogen[2], the obvious alternative is to operate at temperature above liquid conditions at atmospheric pressure, where the thermodynamic efficiency of the refrigerator increases, thus reducing electrical consumption. We show in Fig. 2 a selection of measured values for the inverse Coefficient of Performance (COP) of cryo-plants, where we have defined the COP as:

---

[1] The cost is estimated based on the HL-LHC magnet production, excluding R&D and tooling, taking the 2019 prizes for superconductor, other materials, consumables and labor and adjusting for a 20 % cumulated inflation in the five years period 2019-2024.
[2] Hydrogen or neon come into question when devising a cooling system in the range of 20 to 30 K. This is beyond the scope of the discussion in this paper, and inessential to the main results.

$$COP = \frac{\dot{Q}_{Cold}}{P_{Hot}} \qquad (2)$$

Above, $\dot{Q}_{Cold}$ is the power removed at the cold end, and $P_{Hot}$ is the electric power required by the cryogenic machine to transfer $\dot{Q}_{Cold}$ from a cold source to a hot sink. The inverse of COP is hence the electric power required to remove a given heat load at the cold end.

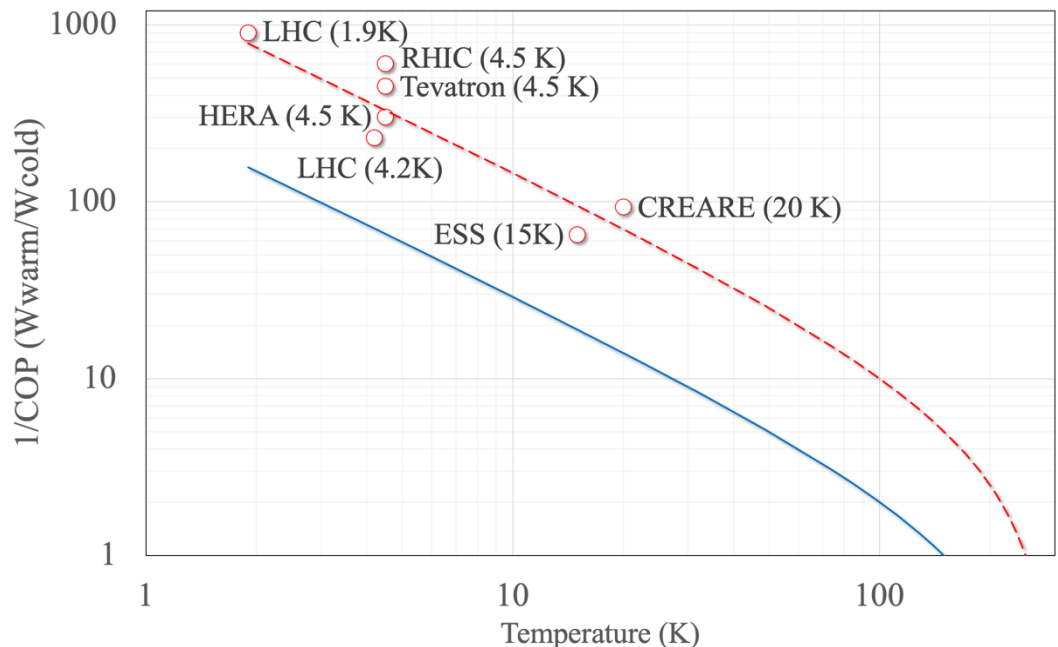

Figure 2. Inverse COP measured in cryogenic machines that have been built and operated in accelerators. The solid blue line is the ideal value for a reversed Carnot cycle (Carnot Refrigerator), the dotted red line is a guide to the measured COP having the same dependence on cold end temperature as the Carnot efficiency.

The points in Fig. 2 correspond to values from cryogenic machines that have been built and operated in accelerators, with cold power ranging from 3 to 30 kW, plotted as a function of the temperature of the cold end. The solid line is the theoretical lower limit provided by the efficiency of a reversed Carnot cycle. The dashed line has the same functional dependency of the Carnot efficiency on temperature and has been adjusted vertically to guide the eyes among the measured points.

We see that the ratio of efficiency going from 4.2 K to 20 K for real machines is broadly in good agreement with the ratio evaluated using an ideal reversed Carnot cycle. The scattering is due both to the difference in the refrigerator capacity among different cryogenic machines [STR-1974], as well as the evolution of technology in the past half century. The summary plot above is a practical confirmation of the potential for a significant gain in energy efficiency, i.e. a factor four if the temperature of the cold end is increased from 4.2 K to the range of 15K to 20 K. A second benefit of an increased operating temperature is the fact that the thermal management of the magnet cold mass becomes easier. Requirements on thermal shields, vacuum levels and assembly tolerances are relaxed, thus facilitating the construction, integration and installation of the magnet systems in their cryostats, and decreasing the unit cost.

As to the helium inventory, the density of helium at 20 K and atmospheric pressure is about 1/50$^{th}$ of the density in liquid conditions, 4.2 K and atmospheric pressure. In reality, operation at higher cryogenic temperature will entail also the need for higher operating pressure, to maintain good heat removal characteristics and limit the loss associated with the circulation of the cryogen. Studies of force flow cooling [BOT-2024-2] have shown that efficient operation of a 10 to 20 K system requires operation at 10 to 20 bar, which is associated with larger helium density and hence inventory. Still the density of helium at 20 K and 20 bar is a factor three lower than that at liquid conditions, 4.2 K and atmospheric pressure. We can therefore expect a reduction factor of the helium inventory of at least three, and higher if the system is designed for minimal cryogen use.

These considerations of energy consumption, helium availability and risk would drive cryogenics towards operating range in the vicinity of 20 K, where, in addition, alternative fluids could be envisaged. It is evident that this could be made possible only using HTS materials, which provides the first strong motivation for inserting this new technology in accelerators.

**Cost of high field superconductors**

The superconducting material cost is one of the main drivers in high field magnets. The reason is twofold:

- High field superconductors, LTS and HTS tend to be expensive because of the price of raw materials, the complexity of the process, and the relatively small mass of material produced per year, hence lesser opportunity for process optimization. We expand on this point in this section;
- More superconducting material is needed to achieve higher fields. This is because of the combination of the drop of critical current density with field, as well as the need to satisfy mechanical and protection limits that become more stringent as the field increases and results in the natural tendency to reduce current density. This is the subject of the next section.

As shown by [COO-2005], the cost of a superconductor can be obtained from the cost of raw materials, applying a multiplier P that depends on the complexity, maturity and scale of the manufacturing process. The typical value of P for Nb-Ti is 3, indicating a mature and relatively simple process, used for the production in large scale (several thousand tons per year). For $Nb_3Sn$ the present value of P is rather around 6 to 10. $Nb_3Sn$ is produced by an industrial metallurgy process, with many similarities to Nb-Ti. There is therefore potential for some reduction, up to the best case of a factor three projected considering the relatively large production for a next step collider of the size of the FCC-hh [BAL-2015]. The issue is that in spite of the efforts of large scientific projects such as ITER (approximately 600 tons production) and HL-LHC (approximately 30 tons production, including R&D) present market demand and production levels remain too small to trigger industrial production of high-grade $Nb_3Sn$ on a scale comparable to that of Nb-Ti. Besides the technical challenges that still remain to be mastered and translated into an industrial production, even a next step collider such as the FCC-hh only demands a total production in the range of 1000 to 2000 tons/year over three to five years, comparable to the present capacity of a single Nb-Ti producer. Such quantities may not be enough to justify the required investment, especially without a guaranteed market follow-up. It is hence unlikely that the cost of Nb$_3$Sn will decrease significantly with respect to present values. This, by incident, is akin to what has happened in the case of the production of wire for ITER, where the price per kg of material has remained comparable to values at the beginning of the development, in the late 1980's.

For HTS materials, on the other hand, P is much larger. Taking the specific case of REBCO, and using the present range of cost, 150 to 200 USD/kA m, an evaluation of the raw material costs would lead a factor P of 100 to 200, mainly influenced by the cost of Ag. While this the result of the complexity of the process, it also points to the potential for a significant cost reduction through production expansion and scaling [MAT-2012]. This is exactly what was witnessed with the procurements in the last fifteen years, as a result of the scale-up of production capacity mainly driven by the demand originating from R&D on compact thermonuclear fusion devices. We have plotted in Fig. 3 the cost per unit length and unit current carried by the superconductor, for Nb$_3$Sn and HTS, based on CERN orders to industrial manufacturers. HTS orders were mainly of REBCO, for which we refer the unit current capacity to the worst direction (i.e. B parallel to the c-axis). The actual cost has been scaled to yield arbitrary units and avoid disclosing financial offers. We have chosen to compare normalized costs at a field of 12 T and at a temperature of 4.2 K, the typical range where Nb$_3$Sn is expected to have a leading edge. Trendlines are only reported to guide the eye, they are not indicating expected evolution.

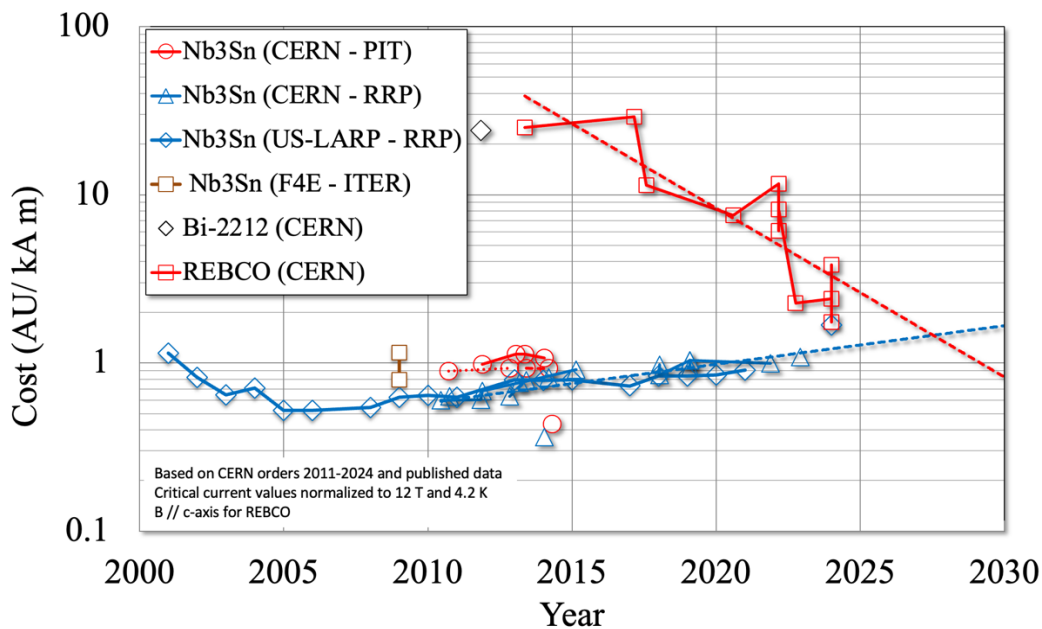

Figure 3. Cost (arbitrary units) per unit length and current transported, for Nb$_3$Sn and HTS (mainly REBCO), derived from purchases at CERN in the last thirteen years. The normalization is done for a field of 12 T and an operating temperature of 4.2 K. Critical current for REBCO is evaluated in the worst orientation, i.e. B parallel to the c-axis. The normalized cost of the procurement of Nb$_3$Sn for US-LARP R&D and US HL-LHC magnets, as well as EU ITER procurement, are reported for information.

It is very suggestive that HTS costs per unit current are approaching, and matching those of Nb$_3$Sn, even today. Had we taken 16 T as the benchmark field, the normalized cost of HTS would be well below that of HEP-grade Nb$_3$Sn. A legitimate question is whether the demand for REBCO will remain and sustain this trend on the medium to long term, which is an issue shared by all high-field superconducting materials, including $Nb_3Sn$. In addition, many issues of magnet science and technology need to be resolved. Still, in spite of the caveats, this analysis clearly points to a great potential for cost reduction using HTS, and provides a second strong motivation for the need for a focused effort in making HTS accelerator magnets a reality.

**High current density in high field accelerator magnets**

We mentioned earlier that the cost of accelerator magnets depends crucially on the amount of superconducting material and tends to increase as the coil cross section grows. We can provide a simple scaling to show this effect by looking at the field produced by a sector coil with current density $J$, free bore $R_{in}$, width $w$ and pole angle $\varphi$:

$$B = \frac{2\mu_0}{\pi} J w \, sin(\varphi) \tag{3}$$

If we fix a given dipole field $B$, the cross section of the corresponding coil A scales like:

$$A_{coil} = 2\varphi(w^2 + 2R_{in}w) \sim \frac{1}{J^n} \tag{4}$$

Above, $n$ is an exponent ranging from the lower limit value of one, obtained for very thin dipole coils of large bore, to the upper limit value of two, for very thick dipole coils of small bore. In the range of dimensions of typical dipoles ($B \approx$ 10…20 T, $R_{in} \approx$ 25…75 mm, $w \approx$ 15…60 mm) it is approximately 1.3.

This scaling clearly demonstrates the interest to increase the current density as much as possible, compatibly with operating margin, mechanics and quench protection. The increase in current density results an over-proportional decrease of the coil cross section and corresponding decrease in the magnet cost. In fact, this has been the natural trend, observed throughout the history of the superconducting collider, as shown in Fig. 4 reporting the engineering (insulated cable) current density $J$ in the dipole magnets from the beginning of the Tevatron to the present High Luminosity LHC magnets. $J$ has nearly doubled since the initial values of the Tevatron and HERA (295 A/mm$^2$) to the present values of the 11T dipoles for HL-LHC (520 A/mm$^2$), thanks to the superior critical current density properties of $Nb_3Sn$ in high-field. Applying the scaling above, the increased current density in the 11T dipoles translates in a reduction by factor 2.5 in the coil cross section that would be required to generate the same dipole field if the values of $J$ of Tevatron were taken. This is an evident gain in cost, which explains the trend. This trend is likely to continue in the future, towards increased performance, but this will require resorting to materials with high critical current density at high field, such as HTS. In fact, this is not the only reason.

To elaborate further, we have also reported in Fig. 4 the values of engineering current density of R&D high-field $Nb_3Sn$ magnets built and tested in the range of 14 to 16 T, FRESCA2 [ROC-2019] and the Racetrack Model magnet, RMM [PER-2015], as well as the design values for the FCC-hh 16 T dipoles based on either cos-theta or blocks coil geometry [ABA-2019-2]. We see a clear deviation from the trend observed in the collider dipoles, with a significant drop of $J$, even in the rather aggressive designs selected for the FCC-hh. One crucial reason is the fact that Nb$_3$Sn performance is limited by the maximum stress that can be applied to the conductor, proportional to the product $J$ x $B$. In practice, values of peak acceptable stress were already reached with the HL-LHC dipole designs. For this magnet technology, hence, an increase in field will correspond forcibly to a decrease in current density. Finally, without expanding further, we recall that quench protection poses comparable challenges and limitations, also leading to the need to decrease the engineering current density as the design field is increased. The corollary is that the coil cross section, and the magnet cost, will increase more than proportionally to the field generated.

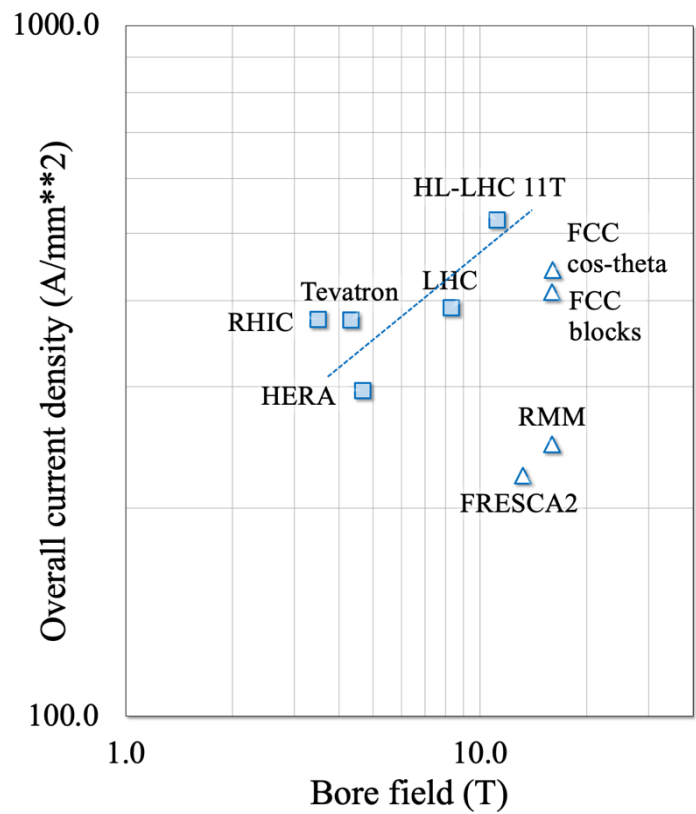

Figure 4. Engineering current density (insulated cable) of the collider dipole magnets built to date. An average of inner and pouter layer has been taken for the LHC. The line is intended only to guide the eye. Also reported for comparison the engineering current density of R&D magnets (Racetrack Model magnet, RMM [PER-2015], and FRESCA2 [ROC-2019]) built and tested, as well as the 16 T FCC dipole designs based on a cos-theta and blocks coil cross section [ABA-2019-2].

An opportunity to control the cost increase associated with the reduction of $J$ is to profit from the renewed interest in magnets with controlled resistance between turns, also generally referred to as "Non-Insulated" (NI) windings. Such windings have been successfully demonstrated in ultra-high-field solenoids, achieving exceptional bore fields of 45.5 T with an engineering current density of 1420 A/mm$^2$ [HAH-2019]. Although these are only small proof of principle demonstrations, such an increase aligns seamlessly to the trend of $J$ increase suggested by Fig. 4.

**Effect of magnet technology on costs**

We can be more specific on the cost expectations and scaling driven by the above arguments. For this study we apply a much-simplified cost model, described in Appendix I, and calibrated to the unit cost of the LHC 8T dipoles, built with Nb-Ti, the HL-LHC 11T dipoles, built with Nb₃Sn, and adjusted to the FCC CDR cost scaling. We do not claim that this model is a precise absolute costing, as many factors need to be considered in a proper cost evaluation (e.g. multiple production sites, learning curves, material cost escalation depending on TRL, etc.). Still, the model should capture well the scaling relation between the CAPEX related to magnet construction cost, and the magnetic field, across different materials and technologies.

We have taken for our analysis the cross section and features of specific magnet designs from published studies, in particular:

- The HL-LHC **11T** LTS dipole, 60 mm bore, built with $Nb_3Sn$, with $J$ of 520 A/mm$^2$ [BOR-2019];
- The FCC **16T** dipole design, 50 mm bore, based on $Nb_3Sn$ graded block coils, graded, with highest $J$ of 500 A/mm$^2$ in the low-field grade [SOR-2017];
- One of the **HE-LHC** 20T dipole designs, 40 mm bore, based on hybrid $Nb_3Sn$ and REBCO coils, with average $J$ of 380 A/mm$^2$ [TOD-2014];
- The **LHC Tripler** 24T dipole design, 40 mm bore, based on hybrid $Nb_3Sn$ and BSCCO coils, with average $J$ of 630[3] A/mm$^2$ [MCI-2005].

All magnets above are double aperture. In addition, we have considered the following scaled designs, namely:

- A **12T** LTS dipole designed as a sector coil using the HL-LHC $Nb_3Sn$ MQXF cable [TOD-2021], 50 mm bore, with a $J$ of 560 A/mm$^2$. This design is one of the present attempts [FOU-2024] to resolve some of the mechanical limits encountered in the HL-LHC $Nb_3Sn$ 11T dipoles, using a wider cable, still producing a cost-effective, albeit lower field option for a future hadron collider;
- Three full HTS (REBCO) dipole designs at increasing field strength dipole obtained scaling the cross section of the FCC 16T dipole (**16T HTS**), of the HE-LHC 20T dipole (**20T HTS**), and the LHC Tripler 24T dipole (**24T HTS**) assuming a $J$ of 1000 A/mm$^2$ and operating at 20 K. The reason for this choice is to profit from the recent advances in REBCO manufacturing for fusion, which has demonstrated routine performance at this level of engineering current density at 20 K [MOL-2021]. Such value also aligns with the past progression of $J$ reported in Fig. 4, justified by the need to make magnets compact. These designs are at the extreme of the evolution of high field dipoles, pushing current density in a region where HTS still has operating margin, but beyond the reach of present technology in terms of mechanical and protection limits. We will come back later to these issues.

The designs were costed using the cost model, under the assumption of "present" cost for both $Nb_3Sn$ and REBCO, as well as using an "aspirational" cost corresponding to a reduction of present cost of the superconducting material by a factor three. This is the factor that was postulated when setting the targets for Nb$_3$Sn development for the FCC-hh [BAL-2015]. This target has not materialized yet, the present situation being rather the opposite, as discussed earlier and demonstrated in Fig. 3. The same factor is anticipated as a likely evolution of the cost of REBCO over the next few years, see again Fig. 3, induced by the expansion and up-scaling of the production capacity at several industrial manufacturers worldwide [MOL-2023]. The main result of the study is shown in Fig. 5, where we plot the magnet cost per unit length as a function of the bore field.

---

[3] The value of engineering current density is not available in literature. It was recomputed computed using the coil cross section and the equivalent thickness of a sector coil that would produce the 24 T dipole field in the bore.

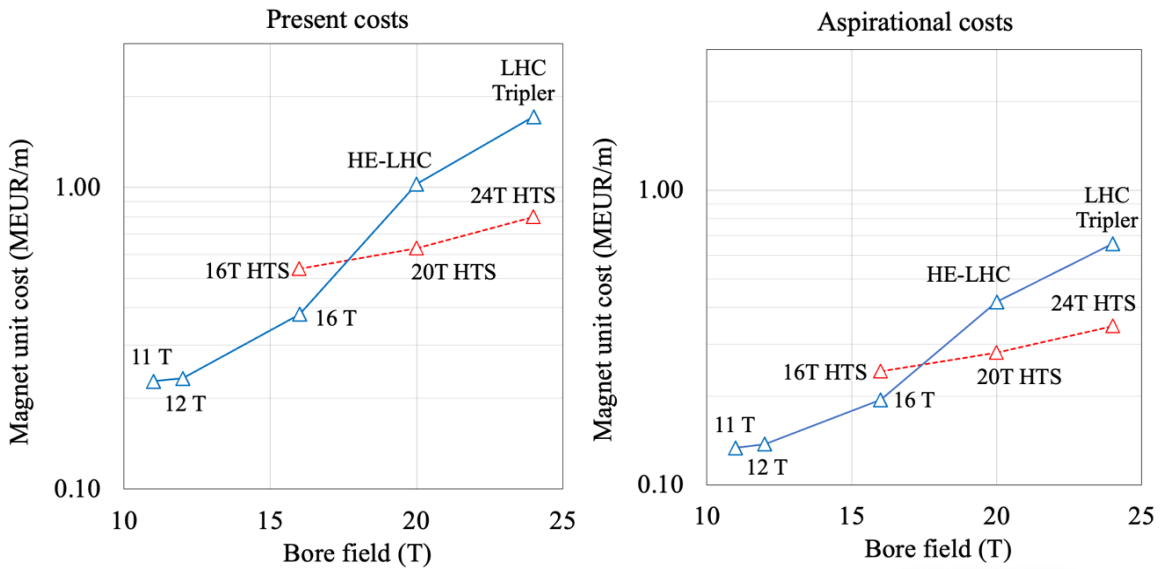

Figure 5. Magnet cost per unit length for a collection of LTS and hybrid LTS/HTS dipole designs (blue symbols and line) and extrapolated HTS dipole designs (red symbols and dotted line) at "present" material cost (left), and at "aspirational" cost (right), corresponding to a reduction of superconductor cost by a factor three.

A first consideration is on the absolute values of the cost figures. Recalling the caveat on the absolute estimates, we can nonetheless compare the values obtained in Fig.5 to those of Fig. 1. As we discussed earlier, the scaling of the cost of technical infrastructure of a next step collider in the range of 10 TeV pCOM energy is expected to drop to 1 MEUR/GeV, which translates to about 130 kEUR/m for the FCC-hh. Assuming for simplicity that the dominant cost is represented by the dipole magnets, we still see in Fig. 5 that present designs fall short of this target by a factor three. This is true even when taking a cost reduction of a factor three ("aspirational" cost), which is possibly realistic for REBCO, but overly optimistic for Nb$_3$Sn. The only range of field that comes close to a suitable cost per unit length is 11 T to 12 T.

The second remarkable feature is the fact that even at present REBCO cost an all-HTS dipole can provide design solutions that are interesting from the point of view of acceptable CAPEX reduction. A 16T HTS dipole for an FCC-hh would be about 30 % more expensive than the Nb$_3$Sn counterpart, but would be able to operate in a temperature range of 10 to 20 K. Beyond this field, an all-HTS solution has clear advantage over previous designs (e.g. LTS/HTS hybrid), and brings the additional benefit of releasing the constraint on operation at liquid helium temperature. As discussed earlier, operating at higher cryogenic temperature improves the energy balance and OPEX.

Finally, if the cost of REBCO will decrease by a factor three with respect to present values, as expected by present analysis [MOL-2023], this will make HTS the material of choice for affordable and energy efficient accelerator dipole magnets in the range of 13 T to 20 T. In this range, Nb$_3$Sn at present cost no longer has an advantage. Even assuming an optimistic and unlikely price reduction for Nb$_3$Sn, the tipping point may shift to 16 T to 18 T, but HTS still has the additional benefit of increased operating temperature, which is not accessible to Nb$_3$Sn. In summary, considerations of performance vs. both capital and operation expenditure (CAPEX and OPEX) underline the disruptive potential of HTS magnet technology for a future circular collider.

**Technical challenges**

So far, we have focused the discussion on various aspects that influence CAPEX and OPEX of the superconducting magnets of a future collider at the energy frontier, addressing the performance and cost targets that need to be met to reach an affordable range of cost, and ensure sustainable operation of the research infrastructure. We have identified an increase of engineering current density to the range of 1000 A/mm$^2$ as crucial to decrease the cost of future high-field accelerator magnets, and operation at a temperature above liquid helium, in the range of 20 K, as crucial to achieving higher energy efficiency and decrease the risk of the helium supply chain. A question so far unresolved is on the technology required to build such magnets. Three main technical challenges need to be addressed:

- Operating margins, in current and temperature, that are customarily required to reach and maintain operating conditions with minimal or no quench training, retain training memory through thermal and powering cycles, and avoid degradation. Margins depend on the critical current dependence on field and temperature, as well as operating temperature. LTS accelerator magnets were built and operated with a range of temperature margins from about 1 K (e.g. Tevatron magnets operated at 4.6 to 4.8 K) to about 2 K (e.g. LHC magnets operated at 1.9 K) and up to 4.5 K (e.g. the Nb3Sn HL-LHC magnets operated at 1.9 K). HTS magnets operating at 20 K, on the other hand, will probably experience no training, and will have the ability to reach critical current, thus making the best possible use of the superconductor material. The reason is the exceptional stability margin, orders of magnitude higher than LTS. Operating margins will still be required, e.g. to accommodate for temperature control in gas conditions, but will not be driven by the superconductor critical surface. As present industrial production has shown the ability to meet a target of 1000 A/mm$^2$ engineering current density, such values seem to be within reach for HTS magnet technology;
- High field magnets experience large electro-magnetic forces, generating stress and strain that can reach material limits. High current density in high field implies high stress and strain. If we take a field range of 16 T to 20 T, and a coil current density of 1000 A/mm$^2$, this corresponds to an electromagnetic pressure in the range of 300 MPa to 400 MPa. This range is a factor two to three above the limit of transverse compression for Nb$_3$Sn. HTS REBCO, on the other hand, is deposited on sturdy substrates and has shown the ability to withstand transverse pressures on its wide face in the range of stress quoted above. Of course, this is only one of the stress components acting on the conductor in an accelerator magnet, and stress distribution in a coil leads to inevitable peaking factors. Delamination and shear remain an issue to be resolved, possibly and partly by magnet design. Still the ability to sustain such high transverse stress is a definite asset towards increasing current density;
- Effective quench management, including a fast quench detection and energy dump, is mandatory to avoid stressing the coil by excessive temperature, thermal stresses or voltages. Here again the combination of high current density and high field implies a high stored energy density. In the same range of field of 16 T to 20 T and coil current density of 1000 A/mm$^2$ as considered above, the energy density of 200 MJ/m$^3$ to 300 MJ/m$^3$. This is a factor two to three times larger than typical values for LTS accelerator magnets, including the Nb$_3$Sn magnets built and designed to date, including HL-LHC and FCC-hh. Present technology is not suitable for this range of current density and stored magnetic energy density, a change is required. The very high stability of HTS materials resolves the issue of magnet training, but may make quench detection and protection significantly more difficult than in LTS. This is where no-insulation or controlled-insulation windings, well suited to HTS, may offer a solution, profiting from fast quench propagation features driven by the combination of resistive heating and inductive coupling during quench. While it is not yet clear that windings with no- or controlled-insulation can be used in a ramping accelerator, because of issues of field lag, field quality, and AC loss, the potential improvement is major, and worth exploring.

In addition to the above challenges, accelerator magnets need a reproducible field quality that allows correction, which can be complemented by feed-back based on direct beam measurements. This should be a matter of debate with beam physics, striving to find the balance between strict beam specifications, magnet technology limitations, and modern beam diagnostics and controls.

Some of these challenges were anticipated when forming first ideas of technology driven magnet R&D programs. The EU-based High Field Magnet (HFM) R&D program entered among the priority R&D lines of the EU strategy for particle physics [ESG-2020] and was initiated under the auspices of the European Group of Large Laboratories Directors (EU-LDG) [LDG-2022]. In the US, the US-DOE Magnet Development Program (US-MDP) pursues similar objectives [MDP-2016]. From their genesis, now dating over five years back, both magnet R&D programs were initiated with main focus on Low Temperature Superconductors (LTS), and in particular Nb$_3$Sn [BOT-2021]. None of the above programs is addressing explicitly accelerator magnet design at current densities in the range of 1000 A/mm$^2$, as would be required to significantly reduce the magnet cost. More recently, the magnet R&D for the Muon Collider [BOT-2025] has initiated activities in this direction by putting priority on HTS as baseline technology for all accelerator magnets. Our analysis suggests that this priority shift should be extended to the other running programs.

In order to guide the development, and based on magnet engineering studies, we have collected in Tab. I a first attempt to set performance targets for REBCO tapes to be used in accelerator magnets. Besides the requirement on the current carrying capacity (note it being referenced to the non-Cu fraction of the tape), we have included considerations on allowable stress and strain, relevant to the mechanical design, and transverse resistance, relevant to current transfer for quench protection and joints.

Table I. Key characteristics of REBCO superconductor for HEP applications

|  |  | Target |
|---|---|---|
| Minimum $J_{C\text{non-Cu}}$ (4.2 K, 20 T) | (A/mm$^2$) | 3000 |
| Minimum $J_{C\text{non-Cu}}$ (20 K, 20 T) | (A/mm$^2$) | 1200 |
| Maximum standard deviation $\sigma(J_{C\text{non-Cu}}(4.2\ K, 20\ T))$ | (%) | 5 |
| Minimum unit length | (m) | 1000 |
| Minimum bending radius | (mm) | 5 |
| Allowable non-Cu $\sigma_{\text{longitudinal non-Cu}}$ (4.2 K) | (MPa) | 1000 |
| Allowable compressive $\sigma_{\text{transverse}}$ (4.2 K) | (MPa) | 600 |
| Allowable tensile $\sigma_{\text{transverse}}$ (4.2 K) | (MPa) | 50 |
| Allowable shear $\tau_{\text{transverse}}$ (4.2 K) | (MPa) | 50 |
| Range of allowable $\varepsilon_{\text{longitudinal}}$ | (%) | -0.1…0.5 |
| Internal specific resistance $\rho_{\text{transverse}}$ (77 K) | (nΩ cm$^2$) | 20 |

**Conclusion –HTS for the future of HEP**

It is clear from the previous discussion that HTS may have a major role in addressing the challenges of the next circular collider at the energy frontier.

HTS materials, REBCO in particular, have exceptionally high critical fields, and can carry high currents at field and temperature levels well above those of LTS materials. Their limitations are no longer the critical properties, margin and training. They hence open the way to fields well above the present HL-LHC standard of 12 T, and can achieve the current densities that are required to wind compact coils and contain material cost of high field accelerator magnets. In addition, REBCO is a material of choice for applications in several other fields of scientific and societal applications. Contrary to HEP-grade Nb$_3$Sn, REBCO has the perspective of a sustainable market, and its cost will decrease further.

Operating at temperature significantly higher than liquid helium, HTS materials offer superior energy efficiency and can reduce the dependency on helium, whose supply has price volatility and availability risks. This is key for the sustainability of a research infrastructure at the scale of a future circular collider

Finally, a last and most important reason is that HTS magnet technology has many overlaps with other fields of magnet science. This is a very strong motivator for present and future research and development with relevance to other scientific and societal applications such as:

- magnetically confined thermonuclear fusion;

- high magnetic field science;
- UHF magnets for nuclear magnetic resonance and research in magnetic resonance imaging;
- fast ramped magnets for radiation therapy;
- compact planar and non-planar coils for superconducting motors and generators.

While a future collider at the energy frontier may depend critically on HTS magnet technology, the HTS magnet technology developed for such a collider would greatly impact other fields of scientific and societal applications. It is exactly this *virtual circle* that has fueled superconducting magnet technology for HEP in the past fifty years, and we wish to foster and nourish it further.

## Appendix I - Cost model for accelerator magnets

Below we give the details of the cost model used for our scaling analysis. We estimate the total magnet cost as the sum of the three main steps of accelerator magnet manufacturing:

$$C = (C_{Coils} + C_{ColdMass} + C_{CryoMagnet}) L_{Magnet} \qquad (AI.1)$$

where $L_{Magnet}$ is the magnet length (we take the magnetic length as good indicator) $C_{Coils}$ is the cost per unit length of manufacturing the superconducting coils, $C_{ColdMass}$ is the cost per unit length of manufacturing a cold mass from the coils, and $C_{CryoMagnet}$ is the cost per unit length of manufacturing a cryostated magnet from a cold mass. We exclude any R&D and tooling cost from the valuation, as would apply to an industrial series production.

The cost of the coils includes the cost of the superconductor itself, $C_{SC}$, the cost of making an insulated cable $C_{Cable}$, as well as the cost of labor and consumables for winding and processing such as heat treatment and impregnation for $Nb_3Sn$, $C_{CoilManufacturing}$. We can write:

$$C_{Coils} = C_{SC} + C_{Cable} + C_{CoilManufacturing} \qquad (AI.2)$$

The cost of the superconductor, which tends to be a large fraction of the total cost of the coil, can be computed from the strand mass in the coil per unit length, $M_{Strand}$, and the superconductor cost per kg, $c_{Strand}$, i.e.:

$$C_{SC} = M_{Strand}\, c_{Strand} \qquad (AI.3)$$

The strand mass per unit length can be computed from the coil cross section $A_{Strand}$ and strand density $d_{Strand}$, including a correction for the cabling angle $\theta$:

$$M_{Strand} = d_{Strand}\, A_{Strand} / \cos(\theta) \qquad (AI.4)$$

where we take a cabling angle of 15 degrees. The cost of cabling and insulation can be estimated to be a percentage $f_{Cable}$ of the cost of the superconductor, or:

$$C_{Cable} = f_{Cable}\, C_{SC} \qquad (AI.5).$$

The fraction $f_{Cable}$ is in the range of 10% for Nb-Ti and $Nb_3Sn$ insulated cables. We assume similar values for insulated HTS cables, while for non-insulated HTS this cost does not apply.

The cost of the coil manufacturing $C_{CoilManufacturing}$ depends on the specific technology, and needs to be calibrated against relevant examples, as we do below. It tends to be higher for magnets built with materials requiring heat treatment and impregnation, such as $Nb_3Sn$ and BSCCO.

In the cost of making the cold mass $C_{ColdMass}$ we include all materials other than the superconducting coils (structure, ferromagnetic yoke), and the labor required for the construction:

$$C_{ColdMass} = C_{ColdMassMaterials} + C_{ColdMassManufacturing} \qquad (AI.6)$$

At our level of estimate, the labor required for the cold mass manufacturing does not depend on magnet technologies. But the materials may be scaled with dipole field because as the field increases, the mechanical structures and the yoke also grow in size and mass. We hence postulate:

$$C_{ColdMassMaterials} = B/B^{Ref}\, C_{ColdMassMaterials}^{Ref} \qquad (AI.7)$$

where superscript Ref indicates a suitable reference point (e.g. LHC, see later).

Similarly, the cost of cryostated magnet manufacturing $C_{CryoMagnet}$ is likely independent on both technology and field, and is obtained by suitable benchmarking to a reference magnet construction (e.g. the LHC, see later).

Following [SCH-2017], which established a very similar cost model for the FCC-hh 16 T dipoles, we use the LHC dipole as a benchmark to obtain reference values of the above parameters. From the LHC experience on the Nb-Ti dipoles series production, the total cost per 14.3 m dipole evaluated in 2018 was about 0.9 MEUR [SCH-2017], or 1.13 MEUR in 2024[4]. This corresponds to a cost total cost per unit length of 79.3 kEUR/m. This cost was split by [SCH-2017] in approximately 1/3 for conductor, 1/3 for structural and other materials, and 1/3 for labor. In fact, [ROS-2013] quotes a cost of 20% to 30% for the conductor, i.e. 19.8 kEUR/m to 26.4 kEUR/m in a dipole (two apertures). The SC coils of the LHC dipole have a strand mass per unit length of approximately 56 kg/m, including the correction for the cabling angle. With a cost per kg of Nb-Ti $c_{Strand}$ of approximately 160 EUR/kg, an estimate of present values, the cost of strand in a coil is about 9 kEUR/m. We attribute the difference to the cabling ($f_{Cable}$ = 0.1) insulation and labor costs, including coil winding. This results in an estimate $C_{CoilManufacturing}$ of 9.9 kEUR/m to 16.5 kEUR/m for Nb-Ti coils. The cost of the cryostat was estimated in 2007 at 100 kCHF [PON-2007], which corresponds in 2024 to approximately 115 kEUR for a LHC dipole, or a cost per unit length $C_{CryoMagnet}$ of 8 kEUR/m, where we have assumed that the dominating cost is the cryostat itself. This yields a cost of materials in the cold mass other than the superconductor $C_{ColdMassMaterials}$ of 25 kEUR/m, and a cost of labor $C_{ColdMassManufacturing}$ of 26.4 kEUR/m. We set the reference for the scaling of the cold mass materials taking the value above, i.e a $C_{ColdMassMaterials}^{Ref}$ of 26.4 kEUR/m for a field $B^{Ref}$ = 8.33 T.

For $Nb_3Sn$, we take the construction of the HL-LHC 11T dipoles as a benchmark for suitable adjustment. In this case the cost of the bare superconducting strand can be estimated at 128.5 kEUR/m. This is consistent with the value of the insulated cable of 188 kEUR/m that was declared by CERN to the contractual partner manufacturing coils, GE. The difference with respect to the LHC Nb-Ti dipoles, more than one order of magnitude, is largely due to the higher cost of $Nb_3Sn$ compared to Nb-Ti, a factor 15. The declared value of the coil manufacturing in the CERN-GE contract, labor and consumables, was 148.2 kEUR/m. Also this cost is significantly higher than Nb-Ti LHC dipoles because the manufacturing process of Nb$_3$Sn coils entails additional lengthy and delicate steps such as heat treatment and epoxy impregnation. If we make the assumption that the cost of materials in the cold mass scales with the field, yielding a value of 33 kEUR/m, while the cold mass labor and cryostated magnet costs remain the same as for the LHC dipoles, we arrive at a cost estimate of 404 kEUR/m. This value is consistent with the cost estimate of the HL-LHC Nb$_3$Sn magnet production, once R&D and tooling is removed from the overall project budget.

---

[4] The cumulated inflation index 2017-2024 for the EUROZone can be evaluated at 1.26, https://www.in2013dollars.com/Euro-inflation

Finally, we can benchmark against the cost estimate of the FCC-hh 16 T dipole. In [SCH-2017] the cost of a FCC-hh 16 T dipole was estimated as the sum of conductor material, 670 kEUR, other materials in the cold mass, 600 kEUR, and labor, 400 kEUR. The total cost was evaluated in the range of 1.5 to 2 MEUR per magnet of 14.3 m length, which corresponds to a cost per unit length in the range of 105 kEUR/m to 140 kEUR/m. We can evaluate it independently using the cost algorithm outlined here, using the 2017-2024 escalated cost of materials and labor. Escalating the Nb$_3$Sn cost assumption of 450 EUR/kg, resulting in 567 EUR/kg, scaling the cold mass materials cost as from Eq. (AI.7), resulting in 48 kEUR/m, adding 20 % [SCH-2017] to the coil manufacturing labor costs of the LHC dipoles, resulting in 11.9 kEUR/m, and the cold mass manufacturing labor costs of the LHC dipoles, resulting in 31.7 kEUR/m, and keeping the same cost for cryostat and cryostating, 8 kEUR/m, we obtain a total cost estimate of about 170 kEUR/m, or 2.4 MEUR for a full size dipole. This is a fair match to the estimate of the escalated cost of the dipole provided in [SCH-2017], which in 2024 figures would yield 1.9 MEUR to 2.5 MEUR.

In summary, the cost model outlined here seems to be a good means to compare designs.

Table AI-I. Summary of parameters used for the estimate of cost of superconducting accelerator magnets built with Nb-Ti, Nb$_3$Sn, REBCO and BSCCO.

| | | Nb-Ti | Nb$_3$Sn (present) | Nb$_3$Sn (aspirational) | REBCO (present) | REBCO (aspirational) | BSCCO (present) | BSCCO (aspirational) |
|---|---|---|---|---|---|---|---|---|
| $c_{Strand}$ | (EUR/kg) | 159 | 2274 | 758 | 8013 | 2671 | 17700 | 5900 |
| $d_{Strand}$ | (kg/m$^3$) | 8000 | 8000 | 8000 | 7800 | 7800 | 9000 | 9000 |
| $f_{Cable}$ | (-) | 0.1 | 0.1 | 0.1 | 0 | 0 | 0.1 | 0.1 |
| $C_{CoilManufacturing}$ | (kEUR/m) | 9.9 | 11.9 | 11.9 | 9.9 | 9.9 | 15 | 15 |
| $C_{ColdMassMaterials}^{Ref}$ | (kEUR/m) | 26.4 | | | | | | |
| $B^{Ref}$ | (T) | 8.33 | | | | | | |
| $C_{ColdMassManufacturing}$ | (kEUR/m) | 26.4 | 31.7 | 31.7 | 26.4 | 26.4 | 31.7 | 31.7 |
| $C_{CryoMagnet}$ | (kEUR/m) | 8.0 | | | | | | |

A summary of the parameters used in the cost evaluations is reported in Tab. AI-I. The values of the model parameters assumed for Nb-Ti are directly drawn from the discussion above. For the evaluation of the cost of Nb$_3$Sn magnets we use the present cost of Nb$_3$Sn strands, as well as an *aspirational* value, which is one third of the present value, as declared among the R&D targets in [BAL-2015]. For the Nb$_3$Sn coil manufacturing and cold mass assembly we use the LHC dipole values, each being increased by 20 % to account for the higher complexity of the process [SCH-2017]. For REBCO magnets we take the present cost of tape, converted in cost per unit mass, as well as an *aspirational* value scaled by a factor three, as done for the FCC Nb$_3$Sn strand. The REBCO coil manufacturing and cold mass assembly, on the other hand, are taken identical to those of the LHC Nb-Ti dipoles, as REBCO does not need heat treatment. Finally, for BSCCO magnets we take present and *aspirational* superconductor cost, analogous to what done for other materials. For the BSCCO coil manufacturing we use the LHC dipole values increased by 50 % to account for the additional complexity of a high temperature heat treatment, with controlled over-pressure and partial oxygen atmosphere. For the cold mass, we assume complexity and values identical of those of $Nb_3Sn$ magnets. All other values are identical or scaled from those of the LHC dipoles.